\documentclass[twocolumn,showkeys,aps,prl,showpacs]{revtex4-1}
\UseRawInputEncoding
\usepackage{graphicx}
\usepackage[CJKbookmarks,dvipdfm,colorlinks,linkcolor=blue,citecolor=blue]{hyperref}

\begin{document}

\title{Out-of-plane high-temperature ferromagnetic monolayer CrSCl with large vertical piezoelectric response}

\author{San-Dong Guo$^{1}$, Xiao-Shu Guo$^{1}$, Yu-Tong Zhu$^{1}$ and Yee-Sin Ang$^{2}$}
\affiliation{$^1$School of Electronic Engineering, Xi'an University of Posts and Telecommunications, Xi'an 710121, China}
\affiliation{$^2$Science, Mathematics and Technology (SMT), Singapore University of Technology and Design (SUTD), 8 Somapah Road, Singapore 487372, Singapore}
\begin{abstract}
For two-dimensional (2D) material, piezoelectric ferromagnetism (PFM) with large out-of-plane piezoresponse is highly desirable for multifunctional ultrathin piezoelectric device application.  Here,  we predict that Janus monolayer CrSCl is an out-of-plane ferromagnetic (FM) semiconductor with large vertical piezoelectric response and high Curie temperature. The predicted out-of-plane piezoelectric strain coefficient $d_{31}$ is -1.58 pm/V, which is higher than ones of most 2D materials (compare absolute values of $d_{31}$).
The large out-of-plane piezoelectricity is robust
against electronic correlation and biaxial strain, confirming reliability of large $d_{31}$.  Calculated results show that tensile strain is conducive to high Curie temperature, large magnetic anisotropy energy (MAE) and large  $d_{31}$.
Finally, by  comparing $d_{31}$ of CrYX (Y=S; X=Cl, Br I) and  CrYX (Y=O; X=F, Cl, Br), we conclude that the size of  $d_{31}$ is positively related to electronegativity difference of X and Y atoms. Such findings can  provide
valuable guidelines for designing 2D piezoelectric materials  with large vertical piezoelectric response.

\end{abstract}
\keywords{Ferromagnetism, Piezoelectronics, 2D materials ~~~~~~~~~~~~~~~~~~~~~~~~~~~~~Email:sandongyuwang@163.com}

\maketitle

\section{Introduction}
The multifunctional 2D piezoelectric  materials can give rise to
unprecedented opportunities for intriguing physics, and  provide a potential platform for multi-functional electronic devices\cite{a1}.
The coexistence of piezoelectricity,  electronic topology and ferromagnetism, namely piezoelectric quantum anomalous Hall insulator (PQAHI), has been predicted
in Janus monolayer  $\mathrm{Fe_2IX}$ (X=Cl and Br)\cite{gsd1}, which provides possibility to use the piezoelectric effect to control quantum anomalous Hall (QAH) effects. The  piezoelectric  properties of ferrovalley (FV) materials have been investigated\cite{gsd2,gsd2-1}, and the anomalous valley Hall effect induced  by piezoelectric effect has been proposed in $\mathrm{GdCl_2}$ monolayer\cite{gsd2-1}. Searching for PFMs has been a research hotspot\cite{qt1,q15,q15-0,q15-1,q15-2,q15-3}. An eminent 2D PFM with a typical triangle lattice structure should meet these conditions: (1) the strong FM coupling, which  means high Curie temperature; (2) the out-of-plane  magnetic anisotropy, which means a long-range phase, not a quasi-long-range phase; (3) the large out-of-plane piezoresponse, which is highly desirable for ultrathin piezoelectric device application. However, the reported PFMs satisfy  some of these conditions. For example NiClI monolayer, the large
out-of-plane piezoelectricity ($d_{31}$=1.89 pm/V) has been predicted, but it has very weak FM coupling and in-plane magnetic anisotropy\cite{q15-3}.
For  $\mathrm{InCrTe_3}$ monolayer, it has large FM coupling and out-of-plane  magnetic anisotropy, but the out-of-plane piezoelectricity is  weak ($d_{31}$=0.39 pm/V)\cite{q15-2}.
Therefore,  searching for  2D PFMs with strong out-of-plane FM  coupling and large vertical piezoelectric response is significative and challenging.

\begin{figure}
  \includegraphics[width=7cm]{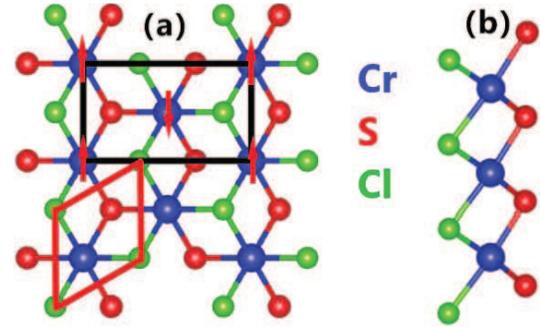}
  \caption{(Color online)For Janus CrSCl monolayer,  (a): top view and (b): side view of  crystal structure. The primitive (rectangle supercell) cell is
   marked by red (black) lines. The red arrows represent the spin direction of Cr atoms.}\label{t0}
\end{figure}

\begin{figure*}
  \includegraphics[width=16cm]{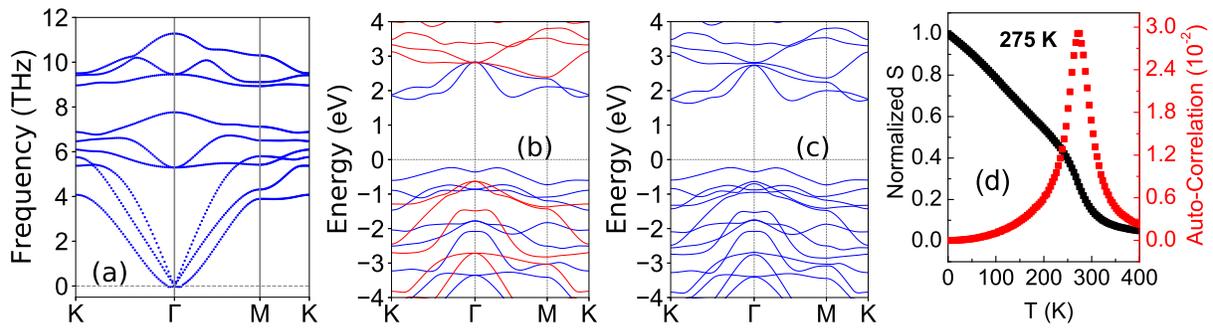}
  \caption{(Color online)For Janus CrSCl monolayer,  (a): the phonon dispersion curves; (b): the energy band structures with GGA; (c): the energy band structures with GGA+SOC; (d): the normalized magnetic moment (S) and auto-correlation  as a function of temperature. In (b),  the blue (red) lines represent the band structure in the spin-up (spin-down) direction.}\label{t1}
\end{figure*}

2D Janus materials  have attracted increasing attention\cite{q16,q16-1}, and the representative MoSSe monolayer has been  successfully fabricated in experiment\cite{e1,e2}.
Due to broken  out-of-plane symmetry, the out-of-plane piezoelectricity can be observed in these Janus materials. Therefore, 2D Janus materials provide potential platform for  searching good  2D PFMs. Recently, Janus monolayer Cr-based dichalcogenide
halides CrYX (Y=S, Se, Te; X=Cl, Br, I) have been predicted by the first-principle calculations, and the  CrSX (X=Cl, Br, I) are FM semiconductors\cite{q17}.  The size of  out-of-plane piezoelectric coefficient may have a positive relation with electronegativity difference of X and Y atoms for Janus MXY materials with M sandwiched between X and Y\cite{gsd3}.

Due to large electronegativity difference of S and Cl atoms, in this work, we detailedly explore the magnetic, electronic and piezoelectric properties of CrSCl. It is found that CrSCl possesses  an out-of-plane magnetization and high Curie temperature. The predicted  $d_{31}$ is -1.58 pm/V, which is higher than ones (less than 1 pm/V) of most 2D materials (compare absolute values of $d_{31}$). More importantly, the large  vertical piezoelectric response is robust against the electronic correlation and biaxial strain, which ensures  high reliability of large $d_{31}$. Moreover, tensile strain can enhance FM coupling, out-of-plane MAE and $d_{31}$, which is  in favour of eminent 2D PFM. Finally, we investigate   $d_{31}$ of CrYX (Y=S; X=Cl, Br I) and  CrYX (Y=O; X=F, Cl, Br), and the calculated results show  that the large  electronegativity difference of X and Y atoms  indeed  can give rise to large $d_{31}$.

\section{Computational detail}
Based on density functional theory (DFT)\cite{1}, the first-principle calculations are carried out by using the projector augmented wave (PAW) method as implemented in Vienna ab initio Simulation Package (VASP)\cite{pv1,pv2,pv3}.  The exchange-correlation  functional is used by adopting generalized gradient approximation  of Perdew, Burke and  Ernzerhof  (GGA-PBE)\cite{pbe}.  To consider  on-site Coulomb correlation of  Cr-3$d$ electrons,
we adopt  $U$$=$2.1 eV\cite{q17} by the
rotationally invariant approach proposed by Dudarev et al\cite{u}, where only the effective
$U$ ($U_{eff}$) based on the difference between the on-site Coulomb interaction
parameter  and exchange parameters  is meaningful. The spin-orbital coupling (SOC) is included to calculate electronic structures and MAE of CrSCl.  To attain reliable results, we set
the energy cut-off of 500 eV, total energy  convergence criterion of  $10^{-8}$ eV and force
convergence criteria  of less than 0.0001 $\mathrm{eV.{\AA}^{-1}}$.  We add a vacuum space of
more than 16 $\mathrm{{\AA}}$ between slabs along the $z$ direction to eliminate the spurious
interactions.
The phonon spectrum with a 5$\times$5$\times$1 supercell is
calculated by using the  Phonopy code\cite{pv5}.
The Curie temperature is
estimated  by Monte Carlo (MC) simulations, as implemented in Mcsolver code\cite{mc}. A 40$\times$40  supercell
with periodic boundary conditions  and   $10^7$ loops are used  to perform the
MC simulation using
Wolff algorithm.
The elastic  stiffness and piezoelectric stress  tensors  ($C_{ij}$ and $e_{ij}$)   are calculated by using strain-stress relationship (SSR) and density functional perturbation theory (DFPT) method\cite{pv6}, respectively.
The first
Brillouin zone (BZ) is sampled by a 21$\times$21$\times$1 k-point grid for electronic structures and $C_{ij}$, and 12$\times$21$\times$1 for FM/antiferromagnetic (AFM) energies and $e_{ij}$.

\section{Main calculated results}
For Janus CrSCl monolayer, as shown in \autoref{t0},  Cr atoms are surrounded by six anions (three S and three Cl) to form a distorted
octahedral structure,  which  exhibits a hexagonal lattice  by Cl-Cr-S
misalignment stacking with a space group of $P3m1$ (No.156).  The difference in atomic size and electronegativity of S and Cl
atoms will result in inequivalent Cr-S and Cr-Cl bonding lengths and  charge distributions, which gives rise to intrinsic polar electric field and out-of-plane piezoelectric response. To confirm the magnetic ground state of CrSCl,
we calculate the total energies of the FM
and  AFM configurations (see \autoref{t0}). Calculated results show that the energy difference between AFM and FM orderings is 183 meV/rectangle supercell. So, the FM configuration
is the ground state, which is related to the
superexchange interaction.  The optimized lattice constant $a$ is  3.474 $\mathrm{{\AA}}$, and the  Cr-Cl-Cr and Cr-S-Cr  bonding angles are $87.85^{\circ}$ and $94.56^{\circ}$. Based on the Goodenough-Kanamori-Anderson
(GKA) rules, FM superexchange interaction
with near $90^{\circ}$ will dominate the interaction between
Cr atoms, which gives rise to the FM coupling\cite{q18,q18-1}.

To  verify the dynamical stability of of CrSCl,  we calculate its
phonon spectra, as shown in  \autoref{t1} (a).  No
negative frequency phonons are observed in the whole BZ, which implies  that CrSCl is dynamically stable.
To further confirm its thermal stability, we  perform  ab-initio molecular dynamics (AIMD) simulations
at 300 K for 7 ps using a 4$\times$4$\times$1 supercell and a time step
of 1 fs.  As shown in FIG.1 of electronic supplementary information (ESI), during the simulation period,  little energy fluctuation is observed and  the structures at the end of the AIMD simulations
show no structural transitions, indicating that CrSCl is stable at a temperature of 300 K.
To determine the mechanical stability of CrSCl,  the linear elastic constants are calculated, and  the 2D  elastic tensor (Voigt notation) with space group $P3m1$  can be reduced into:
\begin{equation}\label{pe1-4}
   C=\left(
    \begin{array}{ccc}
      C_{11} & C_{12} & 0 \\
     C_{12} & C_{11} &0 \\
      0 & 0 & (C_{11}-C_{12})/2 \\
    \end{array}
  \right)
\end{equation}
 Only
two independent elastic constants ($C_{11}$=62.82 $\mathrm{Nm^{-1}}$ and $C_{12}$=13.67 $\mathrm{Nm^{-1}}$) can be observed,  which  meet Born-Huang
criteria of  mechanical stability  ($C_{11}>0$ and  $C_{11}-C_{12}>0$)\cite{ela},  thereby verifying mechanical
stability of CrSCl.

For 2D hexagonal symmetric system with a typical triangle lattice structure, the magnetocrystalline direction determines  type of magnetic phase transition.
The in-plane one (an easy magnetization plane) means that  there is no energetic barrier to the rotation of magnetization in the $xy$ plane, which will produce
 a Berezinskii-Kosterlitz-Thouless magnetic transition to a quasi-long-range
phase\cite{q19,q19-1}. However, the out-of-plane one can give rise to a long-range FM phase. It is difficult to simulate quasi-long-range
phase with in-plane magnetic anisotropy by DFT calculations. In previous studies, the electronic structures are calculated by  assuming magnetocrystalline direction along $x$ axis, which should be different from ones of quasi-long-range
phase. For most 2D systems, the  magnetocrystalline direction has small effects on their electronic properties. However, some essential influences on electronic and topological properties have been observed in some 2D materials\cite{q20,q20-1}. Although the magnetocrystalline direction  of 2D materials can be regulated by external magnetic field,
the needed magnetic field may be very large (For example, the energetic barrier of 1 meV  is equivalent to applying
an external magnetic field of around 5-10 T.). Thus, it is very realistic to search for 2D out-of-plane magnetic materials with a typical triangle lattice structure.  The intrinsic magnetic anisotropy of CrSCl can be  be determined by  MAE. We define MAE as  energy difference
($E_x$-$E_z$), where $E_x$/$E_z$ is the energy per primitive cell when the magnetization is along the $x$/$z$ direction. The positive/negative MAE means out-of-plane/in-plane direction. The calculated MAE is 36 $\mathrm{\mu eV}$,  indicating  that the intrinsic easy axis of CrSCl is out-of-plane. Thus, CrSCl is a 2D long-range FM material.
\begin{figure}
  \includegraphics[width=8cm]{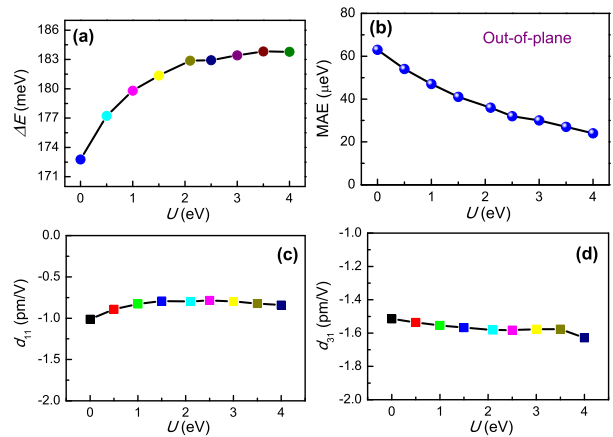}
  \caption{(Color online)For CrSCl monolayer, the energy differences $\Delta E$ between  AFM and FM ordering (a),  MAE (b), $d_{11}$ (c) and $d_{31}$ (d) as a function of $U$.}\label{t2}
\end{figure}

\begin{table*}
\centering \caption{For monolayer  CrYX (Y=S and O; X=F, Cl, Br or I), the lattice constants $a_0$ ($\mathrm{{\AA}}$), the energy difference between AFM and FM ordering $\Delta E$ (meV), the magnetic anisotropy energy MAE ($\mathrm{\mu eV}$),   the elastic constants $C_{ij}$ ($\mathrm{Nm^{-1}}$), the  piezoelectric coefficients   $e_{ij}$  ($10^{-10}$ C/m ) and  $d_{ij}$ (pm/V). }\label{tab0}
  \begin{tabular*}{0.96\textwidth}{@{\extracolsep{\fill}}cccccccccc}
  \hline\hline
 Name &$a_0$&$\Delta E$ & MAE& $C_{11}$&$C_{12}$&$e_{11}$&$e_{31}$&$d_{11}$&$d_{31}$\\\hline\hline
$\mathrm{CrSCl}$ &   3.474  &183      & 36    &62.82    &13.67    &-0.39  &-1.21   &-0.80  &-1.58                           \\\hline
$\mathrm{CrSBr}$ &   3.560  &189     & 52     &61.19    &13.27    &-0.19  &-0.71   &-0.40  &-0.96                       \\\hline
$\mathrm{CrSI}$ &    3.712   &168    & 185    &59.98    &12.72    &0.13   &-0.02   &0.28   &-0.03\\\hline
$\mathrm{CrOF}$ &    3.039         &80    &150     &113.98   &25.19    &-0.21  &-0.97   &-0.24  &-0.70\\\hline
$\mathrm{CrOCl}$ &    3.175         &33    &136     &102.67   &25.14    &1.05   &0.22    &1.35   &0.18\\\hline
$\mathrm{CrOBr}$ &   3.259          &5     &120     &102.13   &26.51    &1.59   &0.79    &2.10   &0.61
\\\hline\hline
\end{tabular*}
\end{table*}

The spin-resolved electronic band structures calculated by using GGA and GGA+SOC are plotted in \autoref{t1} (b) and (c).  Around the Fermi level, the fully
spin-polarized valence band and conduction band are in the same
spin channels. The GGA  results show that  CrSCl is an indirect gap  semiconductor (gap value of 1.98 eV) with both the valence band maximum (VBM) and conduction band minimum (CBM) lying between the K and $\Gamma$ points.  It is noted that the two extremums of top valence band are very close, and the energy difference is only 5.3 meV. With the inclusion of  SOC, the gap is reduced to 1.87 eV. Moreover, sizable Zeeman-type  valley splitting
occurs at the $\Gamma$ point for spin-down direction, and the  splitting is 66.5 meV.  According to the Cr-$d$ band structure projection (FIG.2 of ESI), the Cr atoms change from 4$s^1$3$d^5$ to 4$s^0$3$d^3$ in the electronic arrangement during the synthesis of CrSCl, which means that  CrSCl would have  a theoretical magnetic
moment of 3 $\mu_B$. Calculated results show a net magnetic moment of 3 $\mu_B$ per CrSCl
chemical formula.

To simply estimate Curie temperature $T_C$,  the spin Hamiltonian under
the Heisenberg model can be expressed as:
\begin{equation}\label{pe0-1-1}
H=-J\sum_{i,j}S_i\cdot S_j-A\sum_i(S_i^z)^2
 \end{equation}
where   $J$, $S$ and $A$ are  the nearest exchange parameter,  spin quantum number and  MAE, respectively.
By comparing energies  of  AFM  ($E_{AFM}$) and FM ($E_{FM}$)
 configurations of CrClBr with rectangle supercell, the $J$ with normalized spin vector ($|S|$=1) is determined from these equations:
  \begin{equation}\label{pe0-1-2}
E_{AFM}=E_0+2J-2A~~~~~~~E_{FM}=E_0-6J-2A
 \end{equation}
  \begin{equation}\label{pe0-1-3}
J=\frac{E_{AFM}-E_{FM}}{8}
 \end{equation}
 where $E_0$ is the energy without magnetic coupling. The  normalized magnetic moment and auto-correlation  as a function of  temperature with $J$=22.86 meV are plotted in \autoref{t1} (d), and the predicted $T_C$ is about 275 K.

\begin{figure}
  \includegraphics[width=8cm]{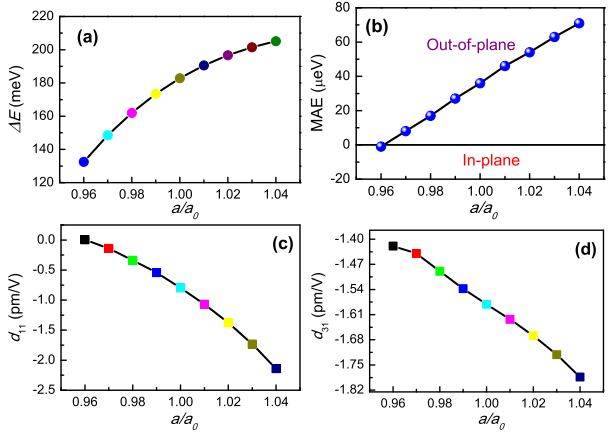}
  \caption{(Color online)For CrSCl monolayer, the energy differences $\Delta E$ between  AFM and FM ordering (a),  MAE (b), $d_{11}$ (c) and $d_{31}$ (d) as a function of $a/a_0$ at $U$$=$2.10 eV.}\label{t3}
\end{figure}

Due to  unique Janus
structure of CrSCl, both in-plane and  out-of-plane piezoelectric response can be observed.  By using  Voigt notation,  the 2D  piezoelectric stress and  strain  tensors for $P3m1$ symmetry can be reduced into\cite{q7,q7-2}:
 \begin{equation}\label{pe1-1}
 e=\left(
    \begin{array}{ccc}
      e_{11} & -e_{11} & 0 \\
     0 & 0 & -e_{11} \\
      e_{31} & e_{31} & 0 \\
    \end{array}
  \right)
    \end{equation}

  \begin{equation}\label{pe1-2}
  d= \left(
    \begin{array}{ccc}
      d_{11} & -d_{11} & 0 \\
      0 & 0 & -2d_{11} \\
      d_{31} & d_{31} &0 \\
    \end{array}
  \right)
\end{equation}
  When  an  uniaxial in-plane strain is imposed,   both in-plane and  out-of-plane piezoelectric response can be induced ($e_{11}$/$d_{11}$$\neq$0 and $e_{31}$/$d_{31}$$\neq$0). However,  by applying  a biaxial in-plane strain, only out-of-plane piezoelectric response can exist ($e_{11}$/$d_{11}$=0, but $e_{31}$/$d_{31}$$\neq$0). These mean that pure out-of-plane piezoelectric response can be achieved by imposed biaxial strain.  Here, the two independent $d_{11}$ and $d_{31}$ can be derived by $e_{ik}=d_{ij}C_{jk}$:
\begin{equation}\label{pe2}
    d_{11}=\frac{e_{11}}{C_{11}-C_{12}}~~~and~~~d_{31}=\frac{e_{31}}{C_{11}+C_{12}}
\end{equation}

We use the  orthorhombic supercell (see  \autoref{t0})  to calculate the  $e_{11}$/$e_{31}$ of CrClS. The calculated $e_{11}$/$e_{31}$ is -0.39$\times$$10^{-10}$/-1.21$\times$$10^{-10}$ C/m  with ionic part 0.29$\times$$10^{-10}$/0.18$\times$$10^{-10}$ C/m  and electronic part -0.68$\times$$10^{-10}$/-1.39$\times$$10^{-10}$ C/m. For both  $e_{11}$ and $e_{31}$, the electronic and ionic contributions  have  opposite signs, and   the electronic part dominates the  piezoelectricity. And then, the  $d_{11}$/$d_{31}$ of CrSCl can be attained from \autoref{pe2}, and the corresponding value is -0.80/-1.58 pm/V.
For most 2D materials, their out-of-plane piezoelectric response is less than 1 pm/V, such as  oxygen functionalized MXenes (0.40-0.78 pm/V)\cite{q9},  Janus TMD monolayers (0.03 pm/V)\cite{q7},
functionalized h-BN (0.13 pm/V)\cite{o1}, kalium decorated graphene (0.3
pm/V)\cite{o2}, Janus group-III materials (0.46 pm/V)\cite{q7-6-1}, Janus BiTeI/SbTeI  monolayer (0.37-0.66 pm/V)\cite{o3}, $\alpha$-$\mathrm{In_2Se_3}$
(0.415 pm/V)\cite{o4} and MoSO (0.7 pm/V)\cite{re-11}. For the needs of practical application, a large out-of-plane piezoelectric response is
highly desired to be compatible with the
nowadays bottom/top gate technologies.  Some progresses have been made for large out-of-plane piezoelectric response, such as NiClI (1.89 pm/V)\cite{q15-3}, TePtS/TePtSe (2.4-2.9 pm/V)\cite{re-6} and $\mathrm{CrBr_{1.5}I_{1.5}}$ (1.138 pm/V)\cite{q15-1}. However, compared with  these materials,  the CrSCl possesses out-of-plane FM ordering with high Curie temperature, which is beneficial to practical application. So, CrSCl is a potential  PFM with large out-of-plane piezoelectric response.

To confirm reliability of large $d_{31}$, electronic correlation is considered to investigate piezoelectric properties of CrSCl.  The lattice constants $a$  at different $U$ (0-4 eV) are optimized, and the change is  about 0.066 $\mathrm{{\AA}}$ with increasing $U$.
It is found that CrSCl is always a FM ground state (see \autoref{t2} (a)), and the $J$ changes from 21.60 meV to 22.97 meV with increasing $U$ (variation of 1.37 meV), indicating that high $T_C$ is robust against $U$. Moreover, the CrSCl has a steady out-of-plane  magnetic anisotropy with increasing $U$ (see \autoref{t2} (b)), and the MAE decreases. The evolutions of energy  band
structures as a function of $U$ are calculated by using GGA+SOC, and the gap versus $U$ is plotted in FIG.3 of ESI.  The CrSCl is always a semiconductor in considered $U$ range, and the gap increases with increasing $U$.
The elastic constants  ($C_{11}$, $C_{12}$, $C_{11}$-$C_{12}$ and $C_{11}$+$C_{12}$) and piezoelectric  stress  coefficients  ($e_{11}$ and $e_{31}$) along  the ionic  and electronic contributions as a function of $U$ are plotted in FIG.4 and FIG.5  of ESI.
It is found that these elastic constants ($C_{11}$-$C_{12}$ and $C_{11}$+$C_{12}$) and piezoelectric  stress  coefficients  ($e_{11}$ and $e_{31}$) have weak dependence on $U$. The piezoelectric  strain  coefficients ($d_{11}$ and $d_{31}$) versus $U$ are plotted  in \autoref{t2} (c) and (d).
It is observed that the $d_{31}$ (absolute value) is always larger than 1.51 pm/V within considered $U$ range.
Thus, the large $d_{31}$ is robust against electronic correlation for CrSCl monolayer.

\begin{figure}
  \includegraphics[width=8cm]{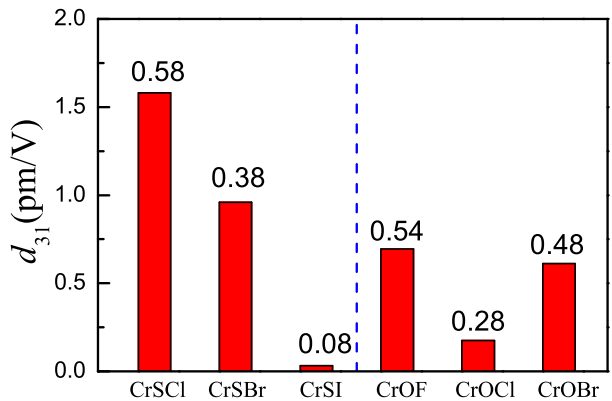}
  \caption{(Color online)For monolayer  CrYX (Y=S and O; X=F, Cl, Br or I), the piezoelectric strain coefficients  $d_{31}$, and the electronegativity difference of X and Y atoms is given in the bar chart.}\label{t4}
\end{figure}

It is well known that the GGA may overestimate lattice constants of materials, and  strain exists naturally in the process of synthesizing materials. So, the strain effects on piezoelectric properties of CrSCl are considered to confirm large $d_{31}$. The $a/a_0$ (0.96-1.04) is used to simulate biaxial strain, where  $a$/$a_0$  denotes   strained/unstrained lattice constant. Taking $U$=2.1 eV, we investigate the strain effects on physical properties of CrSCl.
According to \autoref{t3} (a), all strained CrSCl are  FM ground state in considered strain range. The tensile strain can enhance FM interaction,  which makes for high Curie temperature. The  normalized magnetic moment and auto-correlation  as a function of  temperature at $a/a_0$=1.04 are plotted in FIG.6 of ESI, and the predicted $T_C$ is improved to 303 K. \autoref{t3} (b) shows that tensile strain can enhance MAE, while compressive strain can reduce MAE. At about $a/a_0$=0.96, the easy axis of CrSCl changes from out-of-plane to in-plane. In considered strain range, CrSCl maintains semiconductor characteristics, and the gap as a function of $a/a_0$ is plotted in FIG.7 of ESI. It is found that the gap of CrSCl decreases with $a/a_0$ from 0.96 to 1.04.
The elastic constants  ($C_{11}$, $C_{12}$,  $C_{11}$-$C_{12}$ and $C_{11}$+$C_{12}$) and  piezoelectric  stress  coefficients  ($e_{11}$ and $e_{31}$) along  the ionic  and electronic contributions  as a function of  $a/a_0$ are plotted  in FIG.8 and FIG.9 of ESI. The $d_{11}$/$d_{31}$ versus  $a/a_0$ is plotted in \autoref{t2} (c)/(d). It is found that increasing strain can improve $d_{11}$ (absolute value) due to decreased $C_{11}$-$C_{12}$  and enhanced $e_{11}$ (absolute).  The $d_{31}$ (absolute value) is also enhanced with increasing strain, which is due to reduced $C_{11}$+$C_{12}$.  In considered strain range, the $d_{31}$ (absolute value) is larger than 1.40 pm/V, indicating its robustness against strain. Calculated results show that tensile strain is beneficial to high Curie temperature, large MAE and large $d_{11}$/$d_{31}$ (absolute value).

\section{Conclusion}
In fact, Janus monolayer CrSX (X=Cl, Br I) and  CrOX (X=F, Cl, Br) have all  been predicted to be out-of-plane FM semiconductors\cite{q17,q17-1}, and we  investigate their piezoelectric properties. The optimized lattice constants, energy difference between AFM and FM ordering,  MAE,   elastic constants and  piezoelectric coefficients   $e_{ij}$/$d_{ij}$ are summarized in \autoref{tab0}. They all are mechanically stable due to satisfying   criteria of mechanical stability for calculated elastic constants.  The  piezoelectric  stress  coefficients  ($e_{11}$ and $e_{31}$) along  the ionic/electronic contribution are plotted in FIG.10 of ESI. It is found that,
for both  $e_{11}$ and $e_{31}$, the electronic and ionic contributions  have  opposite signs.  The $d_{31}$ of CrSCl is the largest among CrSX (X=Cl, Br I), while the highest $d_{31}$ among CrOX (X=F, Cl, Br)  is CrOF,  which  may be explained by different  atomic electronegativities of upper and lower layers.
The $d_{31}$  (absolute value) of CrYX (Y=S; X=Cl, Br I) and  CrYX (Y=O; X=F, Cl, Br) along with the electronegativity difference of X and Y atoms are plotted in \autoref{t4}. It is observed that the large  electronegativity difference of X and Y atoms is related with large $d_{31}$.
For CrSX (X=Cl, Br I), the $d_{31}$ decreases with X from Cl to Br to I, and the $d_{31}$ (absolute value) is only 0.03 pm/V due to very small  electronegativity difference of S and I atoms. For  CrOX (X=F, Cl, Br), the $d_{31}$ decreases, and then increases, when X changes from F to Cl to Br.

In conclusion, the electronic structures, magnetic and piezoelectric
properties of CrSCl  are systematically investigated by first-principles calculations.
Calculated results show that CrSCl is an out-of-plane FM semiconductor with high  Curie temperature.  Most importantly, the findings reveal that CrSCl exhibits large  out-of-plane piezoelectric coefficient $\mid d_{31} \mid$ ($>$1.50 pm/V), which is very larger than those of most known 2D materials.
It is found that out-of-plane FM ground state and large  out-of-plane piezoelectric response are  robust against electronic correlation.
The tensile strain can enhance FM interaction, MAE and $\mid d_{31} \mid$, which  provide
the alternative solution for enhancing  out-of-plane piezoelectric effect.  These comparisons about $d_{31}$  (absolute value) of CrYX (Y=S; X=Cl, Br I) and  CrYX (Y=O; X=F, Cl, Br)  provide  a development guide for  searching  Janus 2D materials with large out-of-plane piezoelectric response.
~~~~\\
~~~~\\
\textbf{SUPPLEMENTARY MATERIAL}
\\
See the supplementary material for AIMD results;  elastic and piezoelectric properties of CrSCl as a function of $U$ or $a/a_0$; $e_{ij}$ of CrYX (Y=S; X=Cl, Br I) and  CrYX (Y=O; X=F, Cl, Br).

~~~~\\
~~~~\\
\textbf{Conflicts of interest}
\\
There are no conflicts to declare.

\begin{acknowledgments}
This work was supported by Natural Science Basis Research Plan
in Shaanxi Province of China (No. 2021JM-456). We are grateful to
Shanxi Supercomputing Center of China, and the calculations were
performed on TianHe-2.
\end{acknowledgments}

\end{document}